\documentclass[12pt,english]{article}
\usepackage[T1]{fontenc}
\usepackage[latin9]{inputenc}
\usepackage[square]{natbib}
\usepackage{csquotes}
\usepackage{setspace}
\usepackage{graphicx}
\usepackage{mathrsfs}
\usepackage{amsmath}
\usepackage{amssymb}
\usepackage{babel}
\usepackage[text={16.5cm,23cm},centering]{geometry}
\expandafter\def\expandafter\quote\expandafter{\quote\small}

\makeatletter

\@ifundefined{date}{}{\date{}}
\makeatother

\begin{document}

\title{\textbf{Nietzsche for physicists}}

\author{Juliano C. S. Neves%
\thanks{nevesjcs@ime.unicamp.br%
}}

\maketitle

\begin{center}
{\it{Universidade Estadual de Campinas (UNICAMP),\\
 Instituto de Matemática, Estatística e Computação Científica, CEP. 13083-859, Campinas, SP, Brazil}}
\end{center}

\vspace{0.5cm}

\begin{abstract}
One of the most important philosophers in history, the German Friedrich Nietzsche, is almost ignored by physicists. This author who declared the death of God in the 19th century was a science enthusiast, especially in the second period of his work. With the aid of the physical concept of force, Nietzsche created his concept of will to power. After thinking about energy conservation, the German philosopher had some inspiration for creating his concept of eternal recurrence. In this article, some influences of physics on Nietzsche are pointed out, and the topicality of his epistemological position---the perspectivism---is discussed. Considering the concept of will to power, I propose that the perspectivism leads to an interpretation where physics and science in general are viewed as a game.
\end{abstract}


\section{Introduction: an obscure philosopher?}
The man who said \enquote{God is dead} (GS \S 108)\footnote{Nietzsche's works are indicated by the initials, with the correspondent sections or aphorisms, established by the critical edition of the complete works edited by Colli and Montinari \citep{Colli-Montinari}. The \textit{Birth of Tragedy} is BT, \textit{Human, all too Human} is HH, \textit{Gay Science} is GS, \textit{Beyond Good and Evil} is BGE, \textit{Ecce Homo} is EH, \textit{Twilight of the Idols} is TI, \textit{On the Genealogy of Morality} is GM, \textit{On Truth and Lying in a Non-Moral Sense} is TL, and the posthumous fragments (or notebooks) are PF, indicated by their numbers and years.} is a popular philosopher well-regarded worldwide. Nietzsche is a strong reference in philosophy, psychology, sociology and the arts. But did Nietzsche have any influence on natural sciences, especially on physics? Among physicists and scientists in general, the thinker who created a philosophy that argues against the Platonism is known as an obscure or irrationalist philosopher. However, contrary to common ideas, although Nietzsche was critical about absolute rationalism, he was not an irrationalist. Nietzsche criticized the hubris of reason (the Socratic rationalism\footnote{Socrates, who according to Nietzsche, is \enquote{the archetype of the theoretical optimist}, had \enquote{the imperturbable belief that thought, as it follows the thread of causality, reaches down into the deepest abysses of being, and that it is capable, not simply of understanding existence, but even of \textit{correcting} it} (BT \S 15).}): the belief that mankind could be guided only by reason. That is, the German philosopher was a hard critic of the Enlightenment\footnote{The philosopher suggests a new Enlightenment in several texts. See, for example, the fragments 25 [296], 26 [298], 27 [79] and 27 [80] of 1884.} (or \textit{Aufklärung} in German): to him, the idea of salvation and redemption by means of reason was an equivocated one. Nietzsche accused the hubris of reason,\footnote{In \textit{On the Genealogy of Morality} one reads: \enquote{\textit{Hubris} today characterizes our whole attitude towards nature, our rape of nature with the help of machines and the completely unscrupulous inventiveness of technicians and engineers} (GM III \S 9).} but the philosopher did not deny the use of reason itself. As we shall see, this is clear from Nietzsche's education (\textit{Bildung}). Among his references, there are several natural philosophers and/or scientists. Nietzsche was a reader of Charles Darwin (at least the Darwinian ideas), Hermann von Helmholtz, Roger Boscovich, and others. He tried to stay up-to-date on scientific debate during the 19th century. Therefore, this philosopher who is also considered a poet (\textit{Thus Spoke Zarathustra} is poetry as well) never denied the importance of science. Of course, his scientific view was different, and Nietzsche thought about science from another point of view---by using his perspectivism.\footnote{I will discuss this cardinal concept in Nietzschean philosophy in Section 4.}

According to research in the Nietzschean philosophy, the author's works are didactically divided into three periods. In the first one, there exists an approximation with Romanticism, Schopenhauer, and the German musician Richard Wagner. In the second one, Nietzsche breaks off his friendship with Wagner and stays away from Romanticism and Schopenhauer's influence. The third part is the period where the Nietzschean philosophy acquires its \enquote{full identity} and originality. Science's influence on Nietzsche is present in all of these periods. However, from the second period onward, this influence is more evident. In a book from this period, \textit{Human, all too Human}, Nietzsche says: \enquote{Optimism, for the purpose of restoration} (HH II, Preface 5). That is, Nietzsche identifies science with optimism (an idea originally proposed in his very first book, \textit{The Birth of Tragedy}, where Socratism is criticized) and emphasizes the beginning of a process of a cure. The philosopher recovered his health with the aid of science. His illness was blamed on Schopenhauer's pessimism and Wagner.

In \textit{On the Genealogy of Morality}, Nietzsche defines the purpose of science in our time, modernity: \enquote{\textit{All} sciences must, from now on, prepare the way for the future work of the philosopher: this work being understood to mean that the philosopher has to solve the \textit{problem of values} and that he has to decide on the \textit{rank order of values}} (GM I \S 17). Therefore, as we can see, the importance of science in Nietzschean philosophy transcends the scientific realm.

As we have already seen, among Nietzsche's influences is the multifaceted Roger Boscovich. In the next section, we shall see how important the concept of force from physics (due to Boscovich) was in developing Nietzsche's concept of will to power (\textit{Wille zur Macht}). From that concept, will to power, Nietzsche builds his cosmological view: the eternal recurrence of the same, as we shall see in Section 3. His epistemological position, the perspectivism, is presented in Section 4 with an application to two problems in modern physics: wave-particle duality and the gravitational phenomenon. In Section 5, I use the concepts of will to power and perspectivism to interpret physics---and science in general---as a game.   

\section{Physics in Nietzsche's main concepts}
The Croatian thinker Roger Boscovich (physicist, mathematician, philosopher, etc.) was a decisive reference for Nietzsche's philosophy. Nietzsche's reading of Boscovich's concept of force was an essential ingredient to construct his famous concept of  will to power. In the 18th century, Boscovich studied body collisions. From his research, the Croatian concluded that matter is a manifestation of forces. According to the physicist and physics historian Max Jammer \citep[p. 178]{Jammer}, for Boscovich \enquote{impenetrability and extension [...] are merely spatial expressions of forces, \enquote{force} is consequently more fundamental than \enquote{matter} [...]}. Nietzsche confirms that idea and writes in \textit{Beyond Good and Evil}: \enquote{Boscovich taught us to renounce belief in the last bit of earth that \textit{did} \enquote{stand still,} the belief in \enquote{matter,} in the \enquote{material,} [...]} (BGE \S 12). As well as Boscovich, Nietzsche emphasizes the concept of force (\textit{Kraft}) to the detriment of matter (\textit{Materie}). The material world is a manifestation of forces, which, in the Nietzschean case, as we shall see, are translated into wills to power. 

For Nietzsche, the physical concept of force was important but such a concept was an empty word. In a posthumous fragment, with a bit of irony, this is clear: \enquote{The triumphant concept of \enquote{force}, with which our physicists have excluded God from the world, needs supplementing: it must be ascribed an inner world which I call \enquote{will to power} [...]} (PF 36 [31] of 1885).\footnote{According to \textit{Nietzsche Source} (http://www.nietzschesource.org), the passage translated in \citet[p. 26]{Fragments}, \enquote{our physicists have created God and the world}, is not correct.} In another fragment the idea is stressed: \enquote{A force we cannot imagine (like the allegedly purely mechanical force of attraction and repulsion) is an empty phrase and must be refused rights of citizenship in \textit{science}} (PF 2 [88] of 1885). The will to power, according to Nietzsche's thought, completes the concept of force.

A will to power is a quantum of power, it is \enquote{characterized by the effect it exerts and the effect it resists [...]. The quantum of power is essentially a will to violate and to defend oneself against being violated. Not self-preservation} (PF 14 [79] of 1888), said the philosopher. In this sense, becoming is considered to be a result of the intention to increase power; it is not considered a result of intentions of \enquote{self-preservation}.\footnote{This is the point where Nietzsche finds his disagreement over Darwinian theory.} Above all, Nietzsche writes, \enquote{everything that happens out of intentions can be reduced to the \textit{intention of increasing power}} (PF 2 [88] of 1885). Therefore, will to power means that everything, whether organic or inorganic, \enquote{wants} to increase its power. Such a quantum of power is neither a metaphysical concept nor a substance, it cannot be confused with a being: \enquote{the will to power not a being, not a becoming, but a \textit{pathos}, is the most elementary fact, and becoming, effecting, is only a result of this...}\footnote{We must be careful about the use of \enquote{fact} in that fragment. As we shall see, Nietzsche denies any fact defended by the positivism. The Greek word \textit{pathos} may be translated into affect as well.} (PF 14 [79] of 1888).   

By using the concept of force, in a famous fragment, Nietzsche says what the world is:
\begin{quote}
And do you know what \enquote{the world} is to me? [...]. This world: a monster of force, without beginning, without end, a fixed, iron quantity of force which grows neither larger nor smaller, [...] a play of forces and force-waves simultaneously one and \enquote{many} [...]---\textit{This world is the will to power---and nothing besides!} (PF 38 [12] of 1885).
\end{quote}
The world as will to power can be viewed as forces struggling for more power. A fragment similar to 2 [88] of 1885 has been found, but in this case it indicates the concept of force: \enquote{All that happens, all movement, all becoming as a determining of relations of degree and force, as a \textit{struggle}...} (PF 9 [91] of 1887). There is no goal for all events, \enquote{for all that happens}, then Nietzsche denies any shadow of teleology as we can see in fragment 36 [15] of 1885:
\begin{quote}
If the world had a goal, it could not fail to have been reached by now. If it had an unintended final state, this too could not fail to have been reached. If it were capable at all of standing still and remaining frozen, of \enquote{being}, if for just one second in all its becoming it had this capacity for \enquote{being}, then in turn all becoming would long since be over and done with, and so would all thinking, all \enquote{mind}. The fact of \enquote{mind} \textit{as a becoming} proves that the world has no goal and no final state and is incapable of being.
\end{quote} 
The fragment above shows Nietzsche's refusal to accept an ultimate goal, and this is the reason for rejecting the idea of heat death of the universe (including the second law of thermodynamics), which was already being debated during his lifetime. 

The world as will to power may be read both in the singular or plural forms.\footnote{See also \cite{Muller} for an abundant discussion on these two forms of facing the will to power.} In the singular one, the world is will to power. There is nothing beyond or \enquote{nothing besides!} There is no metaphysical world. Nietzsche denies a metaphysical world and, such as Spinoza,\footnote{In \citet[part III]{Spinoza} the philosopher criticizes those that have considered \enquote{man in Nature as a kingdom within a kingdom}.} considers nature and mankind as the same thing. Nietzsche, in a sense, naturalizes man. Will to power in the plural means a finiteness  of forces. The natural and the human worlds are manifestations of forces or wills to power.

The importance of the concept of force in Nietzsche, besides the concept of will to power, is essential to his cosmological view, and Nietzschean cosmology is the so-called eternal recurrence of the same.  

\section{The eternal recurrence of the same }
Somehow the eternal recurrence of the same (\textit{die ewige Wiederkunft des Gleichen}) is one the most intriguing concepts in Nietzschean philosophy. In the published works, it appears for the first time in \textit{Gay Science}, a book of 1882, in the section, or aphorism, called \textit{The heaviest weight}:
\begin{quote}
What if some day or night a demon were to steal into your loneliest loneliness and say to you: \enquote{This life as you now live it and have lived it you will have to live once again and innumerable times again; and there will be nothing new in it, but every pain and every joy and every thought and sigh and everything unspeakably small or great in your life must return to you, all in the
same succession and sequence [...]. The eternal hourglass of existence is turned over again and again, and you with it, speck of dust!} Would you not throw yourself down and gnash your teeth and curse the demon who spoke thus? Or have you once experienced a tremendous moment when you would have answered him: \enquote{You are a god, and never have I heard anything more divine.} If this thought gained power over you, as you are it would transform and possibly crush you; the question in each and every thing, \enquote{Do you want this again and innumerable times again?} would lie on your actions as the heaviest weight! [...] (GS \S 341).
\end{quote}
In this view, the eternal recurrence appears to be an ethical thought or a challenge. That is, Nietzsche points out a life experience where each singular moment or \enquote{every thing} must be approved. In life each moment---approving it and confirming it---is necessary to accept the possibility of the repetition of the whole life an infinite number of times, \enquote{all in the same succession and sequence}. For an affirmative person, each moment is accepted as it is. This is the supreme \enquote{yes} to existence, according to Nietzsche. In \textit{Ecce Homo} the philosopher stresses that eternal recurrence is the \enquote{highest possible formula of affirmation} (EH, \textit{Thus Spoke Zarathustra} \S 1). On the other hand, the nihilist, who denies the sensible world or the single world,\footnote{Plato, according to Nietzsche, is considered nihilist because he created the ideal world, the word \enquote{where} the Ideas live. Rejecting the sensible world, Plato formulated the True World against the illusory world (the sensible world). In the same way, Nietzsche accuses Christianity because \enquote{Christianity is Platonism for the \enquote{people}} (BGE, Preface). In \textit{Twilight of the Idols} it is written: \enquote{The true world is gone: which world is left? The illusory one, perhaps?... But no! \textit{we got rid of the illusory world along with the true one!}} (TI, \textit{How the true world finally became a fable} \S 6). In a sense, Nietzsche assumes only one world, this world. Then his philosophy is immanent.} is not able to say \enquote{yes} and confirm the existence. Then the eternal recurrence, in this view, is a necessary condition to overcome the nihilism.\footnote{The nihilism, the \enquote{uncanniest of guests}, presents several consequences. In fragment  2 [127] of 1885, the philosopher shows its consequences on science, politics and arts.}

\subsection{A cosmological interpretation}

From another point of view, eternal recurrence is a cosmology or a cosmological interpretation.\footnote{See also \cite{Krueger}, \cite{Nehamas}, \cite{Marton} and \cite{D'Iorio} for discussions on the cosmological meaning of this Nietzschean concept. In  \cite{Neves1,Neves2}, one presents this discussion from our state of the art in cosmology. \cite{Neves1} discusses the possibility of eternal recurrence by means of the scientific knowledge today. Nietzsche himself said that the eternal recurrence \enquote{is the most scientific of all possible hypotheses} (PF 5 [71] of 1886).} The Nietzschean ingredients for this cosmological point of view are: (1) the forces are both finite and conserved and (2) time is infinite. Translating to the language of physics, the first one is indicated by the finiteness of energy in the observable universe. Moreover, Nietzsche considers force a conserved quantity.\footnote{Indeed, a mechanical system described only by conservative forces has its mechanical energy conserved.} To him, this is confirmed by the first law of thermodynamics.\footnote{As we have already seen, the philosopher was a critic of the second law of thermodynamics, but the first law was welcomed by him.} The philosopher wrote about this law and its relation to eternal recurrence: \enquote{The principle of the conservation of energy demands \textit{eternal recurrence}} (PF 5 [54] of 1886). The second one is the eternity of world. For Nietzsche, the recurrence of \enquote{all in the same succession and sequence} is possible with eternity and both conserved and finite forces. All force configurations, within the eternity, according to Nietzsche, would repeat their states. In a sense, Nietzsche works in the same direction of Poincaré,\footnote{A historical description of the Nietzschean eternal recurrence and its similarity to the Poincaré's theorem is found in \citet[vol. II, p. 628]{Brush}. This similarity is stressed in \cite{D'Iorio} as well.} who stated the \enquote{eternal recurrence theorem} years after the German philosopher to begin his first thoughts on the physicality of his concept. 

Contrary to the ethical version, the eternal recurrence of the same, as a scientific thought, appears mainly in the posthumous fragments. One of the most important is the fragment 14 [188] of 1888, called \textit{The new world-conception}, where Nietzsche writes\footnote{This fragment was translated by me directly from critical edition \cite{Colli-Montinari}.}:
\begin{quote}
If the world may be thought of as a certain quantity of forces and as a certain number of centers of force---and every other representation remains indefinite and therefore unusable---thus it follows that, in the great dice game of existence, it must pass through a calculable number of combinations. In infinite time, every possible combination would be sometime reached one time; even more, it would be reached infinite times.   
\end{quote}
As we can see, the two ingredients are present. The first one is indicated by \enquote{certain quantity of forces}, and the second one can be read directly.

The attempts of \enquote{proving} the eternal recurrence by using scientific concepts can be viewed, according to the arguments in \cite{Neves1}, as an expedient used by the philosopher for attracting readers. A scientific form for the eternal recurrence is more acceptable to people immersed in a scientific culture.

\subsection{The possibility of an eternal universe}

Today, cosmology is typically Einsteinian. From solutions of Einstein or Einstein-type equations, cosmological models have been constructed. One of the most important features in these cosmological solutions is the problem of the initial singularity. In the cosmological standard model ($\Lambda$-CDM model),\footnote{CDM means Cold Dark Matter, which is a type of non-relativistic matter able to interact only by means of the gravitational interaction. $\Lambda$ is the cosmological constant developed by Einstein. Today the cosmological constant is the \enquote{source} of the cosmic acceleration, according to several models.} the initial singularity is called the Big Bang. It is interpreted as the initial state of the universe, a singular state where physical quantities, like the matter energy density, and geometrical quantities, like the space-time curvature, diverge. They are unbounded at the initial singularity. There exists a common belief that the Big Bang is a breakdown of Einstein's equations, and a complete quantum theory of gravity would solve this problem. However, there are possible solutions of this problem without the complete quantum theory of gravity. Bouncing cosmologies\footnote{See, for example, \cite{Neves4} and the important review of \cite{Novello}.} appear today as a possibility to avoid the initial singularity within the current physics. If we assume violations of energy conditions (and these violations are acceptable since the observation of cosmic accelerated expansion), regular or nonsingular solutions come from Einsteinian gravitation. The energy conditions relate pressure and energy density of the cosmological fluid, and the fluid description is a good \enquote{approximation} to describe the universe's matter content. Such conditions are necessary hypotheses in the singularity theorems. Assuming the energy conditions and geometrical conditions, the singularity theorems show that space-time possesses a singularity or a singular state. That is, in cosmology, for example, it is possible to show that the matter satisfying the energy conditions leads to the Big Bang or the initial singularity. Then, with energy conditions violations, the singularity theorems are not valid, and it is possible to avoid the Big Bang.\footnote{A detailed study on the singularity theorems (the so-called Hawking-Penrose theorems) is found in \citealt[chapter 9]{Wald}.} In this perspective, the singularity is replaced by a regular transition---a bounce---between a contraction phase and an expansion phase (where we live today). There exists the possibility of constructing cyclic cosmologies in such contexts, where the universe passes through successive phases of contraction and expansion.     

The ekpyrotic cosmology \citep{Lehners}, whose name is inspired by the Stoicism, presents a cyclic cosmology. Moreover, this cosmology provides solutions to the typical problems of the standard model (the flatness, isotropy, homogeneity and horizon problems) without the inflationary mechanism from the $\Lambda$-CDM model.\footnote{The inflationary mechanism assumes a quantum field---the inflaton---able to expand exponentially the space-time fabric in the initial phase of the cosmos (see \citealt{Linde} for a review) and to solve the standard model problems.} That is, the inflationary mechanism and its qualities, which appear during the exponential expansion when the universe was young, are replaced by the ekpyrotic phase, a slow contraction cosmic phase before our current expansion phase. This contraction phase defines the ekpyrotic cosmology. During the contraction phase, besides the solved problems of standard cosmology, there exists a generation of quantum fluctuations that are responsible for structure formation (structures such as galaxies). In the $\Lambda$-CDM model, this achievement is due to inflation. Then the initial singularity problem, or the Big Bang problem, as well as the typical standard model problems and the structure formation, may be solved by adopting an alternative cosmological model.\footnote{In black hole physics it is possible to solve the problem of singularities within the Einsteinian context as well (see \citealt{Neves3}). In particular, the singularity inside the black holes is removed by energy violations.}

Contrary to the critics and some Nietzschean scholars, a cyclic cosmological model is possible today even in Einsteinian theory (the ekpyrotic cosmology, whose origin is in the extra dimensions context, may be thought of as an effective theory in four dimensions, described by general relativity). The door is open to a \enquote{new}  point of view,\footnote{A cyclic view of the cosmos is an old idea. Even Nietzsche writes that \enquote{The doctrine of the \enquote{eternal return}, which is to say the unconditional and infinitely repeated cycle of all things --- this is Zarathustra's doctrine, but ultimately it is nothing Heraclitus couldn't have said too. At least the Stoics have traces of it, and they inherited almost all of their fundamental ideas from Heraclitus} (EH, \textit{The Birth of the Tragedy} \S 3).} where the cosmos is viewed as uncreated, i.e., it is immanent and eternal. The strange death of God, emphasized by Nietzsche, has several meanings: one of the most important is related to the question of the cosmos' eternity. The modern rationality may refuse the Creator or the Demiurge of the universe, in such a way forbidding the Big Bang as an instant of creation because, above all, that instant may be viewed as a shadow of the dead God.\footnote{See the aphorism 108 from \textit{Gay Science}, where the philosopher writes: \enquote{God is dead; but given the way people are, there may still for millennia be caves in which they show his shadow. -- And we -- we must still defeat his shadow as well!}. The thesis that the Big Bang may be interpreted as God's shadow is supported in \cite{Neves1}.}

The question of the possibility of recurrence of the \textit{same} is still an open issue because the knowledge of the structure formation (such as galaxies and galaxy clusters), the black holes evaporation in the contraction phase, and the thermodynamic problem (the entropy would increase in each expansion phase) are not totally solved within our science today. As Nietzsche points out in his \textit{Gedankenexperiment}, his idea of eternal recurrence as a thought experiment assumes the eternal repetition of the same states for generating ethical consequences, \enquote{all in the same succession and sequence}.\footnote{There exists a debate on the recurrence: is the recurrence of the same or of the different? I agree with \cite{Krueger}  because only the recurrence of the same would have an impact on ethical issues.}         

\section{Perspectivism as an epistemological position}
The Nietzschean perspectivism,\footnote{There is an intense debate on the Nietzschean perspectivism. See, for example, \cite{Anderson} on truth and objectivity in Nietzsche's perspectivism and the book organized by Babich \citep{Babich}, which possesses several works on the topicality of this philosophical position.} or his epistemological position, is indicated in a frequently cited posthumous fragment of 1886:
\begin{quote}
Against the positivism which halts at phenomena --- \enquote{There are only facts} --- I would say: no, facts are just what there aren't, there are only interpretations. We cannot determine any fact \enquote{in itself}: perhaps it's nonsensical to want to do such a thing. \enquote{Everything is subjective,} you say: but that itself is an \textit{interpretation}, for the \enquote{subject} is not something given but a fiction added on, tucked behind. --- Is it even necessary to posit the interpreter behind the interpretation? Even that is fiction, hypothesis.

Inasmuch as the word \enquote{knowledge} has any meaning at all, the world is knowable: but it is variously \textit{interpretable}; it has no meaning behind it, but countless meanings. \enquote{Perspectivism} (PF 7 [60] of 1886).
\end{quote}
Denying the thing-in-itself,\footnote{\enquote{The \enquote{thing-in-itself} absurd. If I think away all the relationships, all the \enquote{qualities}, all the \enquote{activities} of a thing, then the thing does \textit{not} remain behind} (PF 10 [202] of 1887).} the fact (or the positivism belief), the final truth (because there is no \enquote{being} and becoming has no goal) and any truth behind or beyond the sensible world (there is no metaphysical world), Nietzsche claims the perspectivism. The knowledge is perspectivistic, it is something \textit{human, all too human}. In a sense, Nietzsche follows Kant and points out the dependence of the human conditions (body structure in the Nietzschean case) to generate knowledge. According to Zarathustra's author, even physics is a perspective or an interpretation, as we can read in \textit{Beyond Good and Evil}: \enquote{Now it is beginning to dawn on maybe five or six brains that physics too is only an interpretation and arrangement of the world (according to ourselves! if I may say so) and \textit{not} an explanation of the world} (BGE \S 14). The world with its \enquote{ambiguous character}\footnote{See GS \S 373.} has become infinite, according to the aphorism \textit{Our new \enquote{infinity}}: \enquote{the world has once again become infinite to us: insofar as we cannot reject the possibility \textit{that it includes infinite interpretations}} (GS \S 374). Above all, interpretations do not reveal any fact or something behind or beyond the sensible world. 

In a provocative form, Nietzsche, as a philologist by trade, criticizes the physicists and their notion of law of nature:
\begin{quote}
You must forgive an old philologist like me who cannot help maliciously putting his finger on bad tricks of interpretation: but this \enquote{conformity of nature to law,} which you physicists are so proud of, just as if --- exists only because of your interpretation and bad \enquote{philology.} It is not a matter of fact, not a \enquote{text,} but instead only a naive humanitarian correction and a distortion of meaning that you use in order to comfortably accommodate the democratic instincts of the modern soul! \enquote{Everywhere, equality before the law, --- in this respect, nature is no different and no better off than we are} [...]. But, as I have said, this is interpretation, not text [...] (BGE \S 22).
\end{quote}  
The old philologist shows the historical and temporal feature of knowledge. Our \enquote{fixation} on the laws of nature, according to Nietzsche, is a feature of modernity. Knowledge is created today by assuming concepts, like the concept of isonomia or equality before the law, which are values for us. Once again, it is emphasized, in the above quotation, that scientific knowledge does not reveal a fact or a \enquote{text}.  

Let us use the Nietzschean perspectivism to look at two questions in modern physics, the wave-particle duality and the gravitational phenomenon, and enrich our discussion on this philosophical concept. 

\subsection{Wave or particle?}

Back to modern physics, Nietzschean perspectivism may help us. With the aid of Nietzsche, dichotomies are banned. For example, the wave-particle duality in quantum mechanics. What is the true reality of matter in the quantum mechanics realm? Wave or particle? For Nietzschean philosophy, both or neither! Both because wave and particle are working interpretations, scientific perspectives of the sensible world (and we shall see what the meaning of \enquote{working interpretations is}). Neither because these interpretations do not show facts or the thing-in-itself. That is, for Nietzsche, there is no perfect correspondence between mind and \enquote{reality}. Because the \enquote{reality} such as we know, the \enquote{reality} given by concepts is not a thing-in-itself, it is a product of human interpretations.\footnote{It must be emphasized that Nietzsche was not a solipsism enthusiast.} Nietzsche rejects the naive realism. Moreover, the philosopher rejects the Platonic idealism and the possibility of existing mathematical entities. We find this rejection in a posthumous fragment: \enquote{Mathematics contains descriptions (definitions) and conclusions from definitions. Its objects do not exist. The truth from its conclusions depends on the correctness of the logical thought} (PF 25 [307] of 1884).\footnote{This fragment was translated by me directly from the critical edition \cite{Colli-Montinari}.} Mathematics is grounded on the error, i.e., \enquote{the invention of the laws of numbers was made on the basis of the error, dominant even from the earliest times, that there are identical things} (HH I \S 19). Denying the identity,\footnote{\enquote{The predominant disposition, however, to treat the similar as identical -- an illogical disposition, for there is nothing identical as such -- is what first supplied all the foundations for logic} (GS \S 111).} such as Heraclitus, and the basis of the classical logic, Nietzsche indicates that even mathematics is a human creation. Hence, Nietzschean philosophy \enquote{solves}, for example, the debate on the reality of the wave function in quantum mechanics. The wave function is only a tool to interpret (itself is an interpretation!).      

\subsection{Force or space-time curvature?}

Another problem in modern physics: what is the true nature of gravity? Is gravity expressed by force or space-time curvature? Is the Einsteinian theory (or something else in the future) the true or the final answer to the gravitational problem? According to Nietzsche, the final answer is only an illusion. Nietzsche denies a final knowledge or a final truth (and even his point of view is an interpretation, a provisional perspective).\footnote{\enquote{Granted, this is only an interpretation too -- and you will be eager enough to make this objection? -- well then, so much the better} (BGE \S 22).} An absolute point of view is absurd and contains a contradiction in terms\footnote{See BGE \S 16.} because every perspective is provisional, temporary. In this sense, both gravitational theories (Newtonian and Einsteinian) are true during some period. Within Nietzschean philosophy, truth, in general, has a polemical definition given by an early text of 1873, \textit{On Truth and Lying in a Non-Moral Sense}. The philosopher asks \textit{what is truth?} and answers:
\begin{quote}
[Truth is] a mobile army of metaphors, metonymies, anthropomorphisms, in short a sum of human relations which have been subjected to poetic and rhetorical intensification, translation, and decoration, and which, after they have been in use for a long time, strike a people as firmly established, canonical, and binding; truths are illusions of which we have forgotten that they are illusions, metaphors which have become worn by frequent use and have lost all sensuous vigour, coins which, having lost
their stamp, are now regarded as metal and no longer as coins (TL \S 1).
\end{quote}
The classical philologist (a young philologist when he wrote that text) indicates in the above quotation truth as \enquote{human relations} or as perspective, as it will be said years later. Specifically, in Nietzschean philosophy, the scientific truth only means: it works for the purposes of subsistence of the type (the scientist is one of several types) and obeys specific rules, which, in the physics case, are both mathematical and empirical. Both rules are interpretations. The first one we have already seen. The second one is stressed in modernity, because both the divine and the metaphysical criteria of truth are rejected. Above all, the empirical obligation (and the scientificity) is motivated by the will to truth (\textit{Wille zur Wahrheit}). According to Nietzsche, the will to truth is grounded on the morality, because the scientist---by assuming the empirical obligation---says: \enquote{I will not deceive, not even myself} (GS \S 344).  Even without God (because \enquote{God is dead} in modernity) and the metaphysical world (the True World is a fable), the will to truth remains a dominant impulse that seeks stability, identity. In our scientific time it appears directly related to the sensible world, i.e., the will to truth seeks to obtain what it wants in our single world: the truth as something that does not change.\footnote{Plato, in \textit{The Republic} (VI, 485b), presents the philosopher's nature and his love for truth. Truth is indicated \enquote{as reality which always is, and which is not driven this way and that by becoming and ceasing to be}. This is a common position even today, and, according to this position, truth is revealed by science because the true scientific theories work independently of time. However, the geocentric model worked during past centuries but it is ruled out today.} It is an error, according to Nietzsche, because the identity, or something that does not suffer corruption, was rejected such as the metaphysical world. In a sense, the scientific work, by using the Nietzschean philosophy, should assume another position: it should look at the truths with new eyes, considering them interpretations, as something temporary, above all, as something \textit{human, all too human}. 

Lastly, the purpose of subsistence of the type is due to knowledge to be similar to \enquote{food} in Nietzsche. A kind of food for the spirit (\textit{Geist}, without any metaphysical sense), which is metaphorically comparable to a stomach: \enquote{[...] \enquote{spirit} resembles a stomach more than anything} (BGE \S 230). After all, the scientific type uses science as food to increase his power.  

The possibility of several interpretations or perspectives is  welcomed in Nietzschean philosophy. In \textit{On the Genealogy of Morality}, Nietzsche says: \enquote{[...] the \textit{more} eyes, various eyes we are able to use for the same thing, the more complete will be our \enquote{concept} of the thing, our \enquote{objectivity}} (GM III \S 12). In this sense, \enquote{objectivity} in Nietzschean philosophy means to have several perspectives on the same thing. Each perspective is a manifestation of impulses or wills to power. During the historical period, mankind lives with/within several perspectives or \enquote{truths}. The Einsteinian and Newtonian theories are \enquote{true}. Of course, the Einsteinian theory contains further elements and it is more sophisticated than the Newtonian theory. Then, it is more \enquote{objective} (such as the dual aspect of the matter in quantum mechanics brings us a more \enquote{objective} look). However, as well as any theory, it is still an interpretation and presents an increase of power to the men who created/supported it. In Nietzschean philosophy, each perspective reflects the body plurality of human body. That is, \enquote{our body is, after all, only a society constructed out of many souls} (BGE \S 19). In his immanent philosophy, soul means impulses or wills to power. Nietzsche has a plural vision, a perspectivistic view on \enquote{reality}.         

\section{Physics or science as a game}
The world as a game may be read in several parts of Nietzsche's works. As early as 1872, the young philosopher wrote the text \textit{Homer's Contest}. The text indicates an aspect that remains unaltered in his mature works: the concept of  \textit{agon}. The Greek concept of \textit{agon} indicates contest, dispute, or struggle. For the mature Nietzsche, both becoming and the \textit{agon} are subsumed under the concept of will to power. The world as will to power means the world as a game or play as well, as we can read in the cited fragment 38 [12] of 1885: the world \enquote{as a play of forces and force-waves [...]}. These are the ingredients of his Dionysian world view. Nietzsche saw in Heraclitus a company, a thinker who claimed a similar point of view: \enquote{The affirmation of passing away \textit{and destruction} that is crucial for a Dionysian philosophy, saying yes to opposition and war, \textit{becoming} along with a radical rejection of the very concept of \enquote{being}--- all these are more closely related to me than anything else people have thought so far} (EH, \textit{The Birth of the Tragedy} \S 3). Becoming as a game---both Nietzsche and Heraclitus are in agreement.\footnote{In a fragment attributed to Heraclitus one reads: \enquote{Lifetime is a child at play, moving pieces in a game. Kingship belongs to the child} \cite[p. 71]{Kahn}.} In a sense, this is a view that may be indicated even today, by using our cosmology. In \cite{Neves2}, one shows that Nietzsche's idea of Dionysian cosmology, with the concepts of becoming and struggle, may be approximated to the notions of the cosmological eras, or eras of domination. In cosmology, the space-time fabric---and its dynamics,  i.e., its expansion, contraction or staticity---is determined by Friedmann's equations. Such equations give the dominant term, which drives the space-time dynamics in a specific period. Each term is a cosmic fluid component in the equations. First, when the universe was young, the radiation dominated the expansion, then the matter. In our present time, the dark energy begins to dominate the cosmic expansion. This picture, somehow, indicates a game or struggle among the matter-energy forms (radiation, matter and dark energy). In each era, an energy-matter form dominates. In a cyclic cosmology, the eras of domination alternate, the sequence radiation-matter-dark energy is repeated, and the \textit{agon} is suggested. The world as a game is a good metaphor from this cosmological perspective as well. 

As a part of the Dionysian world, science is a result of contests as well. The \textit{agon} or contest among scientific perspectives is determinant to the scientific development.  As we have seen, science, in particular physics, obeys rules. Then, such as the Dionysian world, science may be viewed as a game. This conclusion comes from the concepts of will to power and perspectivism: the scientific interpretations struggle for dominance. Science as a game means the most influential game today. From the beginning of modernity, science is the most dominant game. Then, the scientist is a type of player (someone who obtains in science the subsistence of the type) who is immersed in such a sophisticated game and, in general, does not realize that he/she plays it.

\section{Final comments}
Contrary to common belief, Nietzsche was neither an obscure nor an irrationalist thinker. Maybe the reason for this opinion is found in his work. Using aphorisms, Nietzsche created his work differently from the scientific model. Denying all powers to reason, the philosopher pointed out the limitations of reason. However, the German philosopher exhibited an admiration for science and its rationality in his published work and posthumous fragments. With the aid of natural sciences (\textit{Naturwissenschaften}) his concepts were created. In particular, with the aid of physics, Nietzsche developed the concept of will to power. With the Boscovichean concept of force as more fundamental than the concept of matter, the German philosopher thought about the entire world---in terms of forces in struggle. 

His position on the nature of reality is more than relevant today. A philosophy without facts denies a world in itself, or a thing-in-itself. The Nietzschean perspectivism is an epistemological option. \enquote{There are no facts, only interpretations} is the \enquote{fundamental truth} in Nietzschean philosophy. Nietzsche stressed the perspectivistic view of knowledge because \enquote{this world is will to power}, i.e., this world is plural, as well as the interpretations in physics today.      

\enquote{\textit{Long live physics!}},\footnote{See GS \S 335, where Nietzsche talks about the importance of the physicist's honesty.} wrote the philosopher in \textit{Gay Science}. There is no doubt about the physics influence on the Nietzschean main concepts. Will to power and eternal recurrence depend on the physical concept of force to appear. But the contrary is not true. There is a lack of Nietzsche's influence on physics and physicists. However, in our point of view (a perspective), the Nietzschean perspectivism is a good option to interpret the modern results in physics to ban false dichotomies or the improbable final truth.    

From the important concepts of will to power and perspectivism, I derived an interpretation where science is viewed as a game. The Dionysian world reveals a world as contest, a game among wills to power. The multiplicity of perspectives in science, which obey imposed rules, presents science as a game and the scientific activity as \textit{agon}.

\section*{Acknowledgments}
This work was supported by Fundação de Amparo à Pesquisa do Estado de São Paulo (FAPESP), Brazil (Grant
No. 2013/03798-3). I would like to thank IMECC-UNICAMP for the kind hospitality and Elena Senik for reading the manuscript.

\end{document}